\documentclass[12pt,preprint]{aastex}

\newcommand{\alphaFe}{[$\alpha$/Fe]\,}
\newcommand{\FeH}{[\rm Fe/H]\,}
\newcommand{\MgFe}{[\rm Mg/Fe]\,}
\newcommand{\OFe}{[\rm O/Fe]\,}

\def\spose#1{\hbox to 0pt{#1\hss}}
\def\simlt{\mathrel{\spose{\lower 3pt\hbox{$\mathchar"218$}}
     \raise 2.0pt\hbox{$\mathchar"13C$}}}
\def\simgt{\mathrel{\spose{\lower 3pt\hbox{$\mathchar"218$}}
     \raise 2.0pt\hbox{$\mathchar"13E$}}}
\slugcomment{submitted to ApJ}
\usepackage{epsf}

\shorttitle{Chemical Abundances}
\shortauthors{Font et~al.}
\begin{document}

\title{Chemical Abundance Distributions of Galactic Halos and their Satellite 
Systems in a $\Lambda$CDM Universe}

\author{
Andreea~S.~Font\altaffilmark{1},
Kathryn~V.~Johnston\altaffilmark{1},
James~S.~Bullock\altaffilmark{2},
Brant~Robertson\altaffilmark{3}
}

\email{afont@astro.wesleyan.edu, bullock@uci.edu, kvj@astro.wesleyan.edu,
 brobertson@cfa.harvard.edu}

\altaffiltext{1}{Van Vleck Observatory, Wesleyan University, Middletown, 
CT 06459, USA}
\altaffiltext{2}{Department of Physics \& Astronomy, University of California, 
Irvine, CA 92687, USA}
\altaffiltext{3}{Harvard-Smithsonian Center for Astrophysics, 60 Garden Street, 
Cambridge, MA 02138, USA}

\begin{abstract}

We  present  a cosmologically  motivated   model  for   the hierachical
formation of the stellar halo which includes a semi-analytic treatment
of galactic chemical enrichment  coupled to numerical simulations that
track the orbital evolution and tidal disruption of satellites.
A major motivating factor in this investigation is
the observed    systematic   difference  between the   chemical
abundances of stars  in satellite galaxies  and those in the Milky Way
halo.  Specifically, for the  same \FeH values, stars in neighboring
satellite  galaxies  display significantly lower \alphaFe ratios
than stars in the halo.  

We  find that the observed  chemical abundance  patterns are a natural
outcome of the process  of hierarchical assembly  of the Galaxy.  This
result follows  because the stellar halo  in this  context is built up
from satellite galaxies  accreted early on  (more  than $8-9$~Gyr ago)
and enriched in $\alpha$-elements produced  in Type II supernovae.  In
contrast, satellites which  survive today are typically  accreted late
(within the last $4-5$~Gyr)  with nearly solar \alphaFe values as
a result  of contributions from both Type  II and  Type Ia supernovae.
We use our model to investigate the abundance distribution functions
(using  both \FeH and \alphaFe  ratios) for stars  in the  halo and in
surviving  satellites.   Our results  suggest  that the shapes
and peaks in the abundance distribution functions provide a direct 
probe of the accretion histories of galaxies.
\end{abstract}

\section{Introduction}

In   the current framework   of hierarchical structure formation large
galaxies like the Milky  Way are believed to  form through merging  of
many     proto-galactic     fragments   \citep{rees78,searle78,blumenthal84}.   The
observational evidence in support for  this model continues to  mount,
ranging    from   signatures of  different  accretion   events  in the
kinematics of globular clusters and halo stars \citep[see for eg.][for
an extensive review of the  kinematical evidence]{majewski93}, to  the
direct   detection of  satellite  galaxies  in   the  full process  of
disruption  \citep[e.g. the  Sagittarius  galaxy;][]{ibata94,ibata95}.

Chemical abundances are an additional rich reservoir of information which can help uncover the formation of the Galaxy \citep{freeman02}. Stars retain information about the past star formation events and the time of accretion of the halo progenitors from which they came in their chemical abundances even in cases when this information is lost in their phase-space distribution due to phase mixing and violent relaxation. Thus, in reconstructing the history of the Galaxy, chemical data complement and enhance the spatial and kinematic data.

Some chemical abundance data are suggestive of past accretion events. For example, stars with kinematics and chemical abundances distinct from the distribution from which they are drawn --- whether halo \citep{carney96,majewski96}, or disk \citep{helmi05} --- are likely to originate in recently accreted satellite galaxies. \citet{nissen97} find that the abundances of $\alpha$-elements of halo stars decrease with distance in the Galaxy and observe that this may be the result of recent accretions. Also, the kinematics of globular clusters being uncorrelated with their metallicities \citep{searle77} are consistent with the hierarchical scenario. 

Other studies raise some concerns about the validity of the hierarchical structure formation scenario.  Recent observations have revealed systematic differences in the \alphaFe abundances of stars in present day dwarf Spheroidals (dSphs) and stars in the same \FeH  range in the halo.  Specifically, at metallicities of \FeH $\le -1$, stars in the local halo have above solar \alphaFe ratios (\alphaFe $\sim 0.2 - 0.4$), whereas stars in the Local Group dwarf galaxies have nearly solar abundances (\alphaFe $\sim 0.1$) \citep{fuhrmann98,shetrone01,shetrone03,venn04}. The results of these studies show clearly that one cannot assemble an \alphaFe - enriched stellar halo from present day-like satellites which have nearly solar values. The same conclusion can be reached from the study of stellar populations. In trying to relate the young, metal rich populations in the halo with the contribution from present day-like dSphs, \citet{unavane96} estimate that, at the most,  only 6 Fornax-type progenitors or 60 Carinas could have been accreted by the Galaxy. This is in contradiction with the predictions of the hierarchical model, in terms of both the number and the mass spectrum of contributors to a Milky Way-type halo \citep{unavane96,gilmore98}.  Similar discrepancies arise between satellite and halo stellar populations when  comparing the abundances of (intermediate age) giant Carbon stars in each $-$ with the stellar halo apparently less abundant in these stars than current dwarf satellites \citep{vandenbergh94}.
 
However,  it is not obvious that the halo progenitors (or, as they are often referred to, the Galactic ``building blocks'')  were replicas of present day satellite galaxies. As pointed out by \citet{mateo96}, present day dSphs are survivors and therefore special conditions must have precluded them from disruption (late accretion and/or less radial orbital characteristics). These conditions could have allowed present day dSphs to follow a different evolutionary path than the original building blocks. A difference in time of formation alone could explain the lack of intermediate and young populations in the halo relative to the present-day satellites \citep{vandenbergh94,majewski02} ---
running back the clock to the time when the stellar halo was forming, a Carina-type galaxy would have looked much different.

A true test for the hierarchical structure formation needs to use models of satellite galaxy accretions that track their dynamics as well as the development of their stellar populations. Following the chemical evolution of the Galaxy in the full cosmological context is not an easy task, given the current limitations in some of the theoretical areas which need to be covered. In order to tackle this problem one has to have a model that incorporates a realistic merger history of a Milky Way-type galaxy, a detailed follow-up of the dynamical evolution of substructure inside the galactic potential well and a chemical evolution model that includes an accurate treatment of stellar yields, the initial mass function and of  winds from massive and intermediate mass stars. We present here one such attempt of modeling the evolution of the Galaxy. Our approach is a blending of two theoretical methods, one of constructing Milky Way-type stellar halos from accreting substructure \citep{bullock05}, the other of modeling the chemical evolution of individual satellite galaxies by taking into account both the inflow and outflow of matter \citep{robertson05}. Both these models individually suggest solutions to the apparent contradiction between observations and the hierarchical structure formation scenario:  \citet{bullock05} show that halo stars were likely to have been accreted much earlier than the surviving satellite population; and \citet{robertson05}  examine the pattern of chemical abundances in dwarfs with star formation histories truncated at early and late times. This paper extends their study to follow simultaneously both the full phase-space and abundance distributions, allowing us to make definitive statements about the comparison of satellite and halo abundance distributions. Our study follows the phase space distribution of halo and satellite stars with increased numerical resolution than previous studies (see, for example \citet{brook05}), being specifically targeted to these galactic components.

In Section \S 2 we outline the main steps of the methods. Section \S 3 presents the results of mass accretion histories and chemical enrichment of the simulated halos. In \S 4 we compare our results with current observations and in \S 5 we conclude.

\section{Methods}

\subsection{Building up stellar halos through accretion events}

\noindent 
Throughout this work we assume a $\Lambda$CDM  Universe with $\Omega_{m}=0.3$, $\Omega_{\Lambda}=0.7$, $h=0.7$, and $\sigma_{8}=0.9$, in which the Milky Way stellar halo forms entirely from the accretion and disruption of satellite galaxies. Our approach uses a combination of semi-analytic prescriptions  and N-body simulations in order to follow the evolution of the phase-space structure of the stellar halo. A brief summary follows --- for full details, see  \citet{bullock05}. 

Mass accretion histories of twenty Milky Way size galaxies are constructed using the method outlined  in \citet{somerville99} based  on the Extended-Press-Schechter (EPS) formalism  \citep[][]{lc93}. 
We select the eleven histories from this  sample that do not have significant accretions (that is, individual accretions no more than $\sim 10\%$ of the entire mass of the parent halo) during the past 7~Gyr $-$ similar to  the case of the Milky Way \citep{wyse01}. 
For each event in these merger trees an N-body simulation is generated which follows the dynamical evolution of the accreted satellite's dark matter halo from entry into the main halo up to the present time. During the simulation, the satellite orbits in an analytic  disk/bulge/halo potential representing the parent galaxy. The mass and scale of all components of the parent grow with time in proportion to the virial mass and scale of main halo generated by the merger tree.

Each dark matter satellite has a stellar mass that is assigned using a cosmologically-motivated analytic formalism. The infall of cold gas into each satellite is followed from the epoch of reionization to the time of its accretion onto the main halo. The gas mass is translated into a stellar mass through a simple star-formation prescription (see \citet{robertson05} and Section \S 2.2 for more details). Star formation ceases soon after accretion onto the main halo as it is assumed that the gas is lost  at that point due to ram-pressure stripping.  The free parameters in this model are fully constrained by requiring that it approximately reproduces: (i) the number of luminous satellites of the Milky Way; and (ii) the ratio of gas to stellar mass in Local Group field dwarfs (which are assumed to correspond to satellites in the model infalling onto the parent today).

The stellar distribution within each satellite is not modeled self-consistently during the N-body simulations --- rather the stars are "painted on" subsequently by assigning  a variable mass to light ratio to each (equal mass) dark matter particle within a satellite. The stellar profile is assumed to follow a concentration $c=10$ King model \citep{king62}, whose extent is set by the observed trend of core radius with luminosity for Local Group dwarfs.

With all possible freedom in the model now constrained, it additionally produces the observed: stellar mass-circular velocity ($M_{*} - v_{\rm circ}$) relation for Local Group dwarfs (both the amplitude and slope);  distributions of surviving satellites in luminosity, central surface brightness and central velocity dispersion; and stellar halo luminosity and density profile.

\subsection{Chemical evolution of satellite galaxies}

\noindent 
The chemical enrichment of each satellite galaxy is modeled with the chemical evolution code developed by \citet{robertson05}, which tracks the evolution of a range of heavy elements,  in particular of Fe and $\alpha$-elements such as Mg, O, Ca, Ti, etc. This model takes into account the enrichment from both Type II and Type Ia supernovae, as well as the feedback provided by supernovae blow-out and winds from intermediate mass stars. 
Hot gas and metals which are expelled from satellites disperse into the main halo and cannot be later accreted back onto the satellite they originated from. This is a reasonable assumption since the velocity dispersion of the gas heated by SNe ($\sim 200$~km~s$^{-1}$) is generally much larger than that of satellite galaxies (typically, of order $10-50$ km s$^{-1}$ for dwarf galaxies).  

The galactic outflow is assumed to start once the thermal energy of the expelled material (gas or metals) exceeds the gravitational binding energy of the satellite. Since the binding energy per unit mass increases with mass of the galaxy, the wind efficiency is expected to decrease with mass. 
Hence, the galactic wind is modeled  as a function of the depth of the potential well of the satellite galaxy. The normalization of this function is
chosen (in conjunction with a star formation timescale) so that the agreement with the observational constraints described in the previous section is maintained (i.e. the $M_{*} - v_{\rm circ}$ relation; gas to stellar mass ratio of Local Group dwarfs; and the total luminosity of the stellar halo).

It is further assumed that winds are metal enriched,  as metals are shown to escape the potential well with greater efficiency than the gas \citep{maclow99, ferrara00}.  The normalizations of the metal (Fe and $\alpha$-elements) wind efficiencies are treated as parameters. The Fe wind efficiency is constrained by requiring a match to the the stellar mass - metallicity ($M_{*} - Z$) relation \citep{larson74,dekel86,dekel03} over a wide range of dwarf galaxies masses ($M_{*} \sim 10^{6} - 10^{9} \, M_{\odot}$) \citep[see also][and references therein]{mateo98}, while the  $\alpha$-element wind-efficiency is chosen to reproduce the chemical abundance patterns of stars in dSph satellites \citep[][]{venn04}.

Since gas and hence star formation in satellites are truncated upon accretion in our model it  will not reproduce any recent significant star formation and chemical enrichment histories of  the few surviving gas rich systems like LMC and SMC, for which extended chemical modeling may be needed \citep{recchi01}. 
It may also not be applicable to spiral satellite galaxies, in which case it is believed that the regulatory factor in the $M_{*} - Z$ relation it is not the galaxy winds, but a prolonged star formation \citep{tosi98,matteucci01}.
However, it should reasonably represent dSphs, since these systems do not have a significant amount of gas or significant star formation at present day \citep[][]{mateo98,grebel03}.
Although there is evidence for recent star formation in several dSph \citep{gallart99,smecker99}, our model suggests that these systems are likely to have been accreted relatively recently (see \S~3.1), so the majority of their stars are expected to have formed prior to accretion.
Moreover, satellites in general disrupt completely (and hence any continuing star formation would be truncated whether or not gas stripping was efficient) within one or two orbits following their accretion time, so we anticipate our model will reproduce the bulk of halo stars (see \S~4.1 for further discussion of this point).

\subsection{Tracking the combined chemical abundances and phase-space distribution}

\noindent  
Using the methods outlined in \S2.2  we assign each satellite a unique
set of stellar  populations dictated by  its own cosmologically-motivated
mass accretion  history,  which in turn determines its star
formation history and subsequent chemical evolution. We then give each star
particle in  the satellite  a set  of chemical  abundances (  \FeH and
\alphaFe)  drawn at random from  the stellar populations. In doing so,
we assume that all heavy  elements are homogeneously mixed within each
satellite.  This assumption is  rather  simplifying, as heavy elements
may   be  spatially     segregated \citep[see    recent   observations
by][]{tolstoy04}.  However,  the   still limited  information    about
possible metallicity gradients in satellite galaxies precludes us from
adopting   a more  sophisticated  model  for  the metallicity spatial
distributions.

We now have full  phase-space, luminosity and metallicity information for stellar material associated with all the accretion events in our merger trees, and can examine the resulting  spatial and kinematic distribution of metals in our stellar halos and their surviving satellites systems.  Below we present results of a sample of four such simulated galaxy halos, specifically, halos $1-4$ from the 11 simulated halos discussed in \citet{bullock05}. We denote them as halos H1, H2, H3 and H4. Table~1 describes the properties of these four halos such as the total halo luminosity\footnote{Note that the halo luminosities are slightly different from the values quoted by \citet{bullock05}. The differences arise from the different treatment of the star formation rate (i.e., whether feedback is modeled explicitly, as in this paper, or implicitly, as in \citet{bullock05}).}, number of dark and luminous satellites, and the number of satellites which survive until the present day. 

\section{Results}

We describe below the mass accretion histories of the simulated galaxy halos (\S 3.1$-$ 3.2) and the resulting chemical abundances of both surviving satellites and stellar halo (\S 3.3). We then discuss the chemical abundance variations between different halo components as well as between different halo realizations (\S 3.4).

\subsection{Accretion times of Stellar Halos and of Surviving Satellites}
 
An important question in understanding the formation of the Galaxy is whether the stellar halo followed the same pattern as that of the dark matter halo. The hierarchical assembly of galaxy halos is expected to result in a distinctive inside-out growth pattern for the dark matter. Some inferences about the growth of the stellar component of galaxy halos have been made based on scaled-down versions of cluster simulations and  by assuming physically motivated properties for the baryonic material \citep{zhao03,helmi03}.  Recently, simulations targeted specifically for the modeling of stellar galaxy halos confirm that that these too form inside-out, in a short period of time, of a few Gyrs \citep{bullock05}.  

Here we investigate the accretion history of different regions of the stellar halo, with the aim of  identifying which satellites are the dominant contributors and what are their characteristics. This will be important when assessing the build-up of halo chemical abundances, since these retain the signature of the satellite galaxies from which they originate. 

Figure \ref{fig:mass_accr} shows the stellar mass accretion history of each halo binned in time intervals of 0.2~Gyr (here $t_{accr}$ denotes the lookback time of accretion; $t=0$ corresponds to present time). The top row of panels describes the build-up of the entire halo, while the rows immediately below show the build-up of the inner  ($R<20$ kpc),  intermediate ($20$ kpc $<R<$ 50 kpc)  and outer ($50$ kpc $<R<$ 100 kpc) parts of the stellar halo, respectively.  All regions are centered around the solar neighborhood (i.e. $R$ is the distance relative to the position of the Sun at $R_{\odot} =8.5$~kpc from the Galactic center in the disk). Filled histograms represent the stellar material still gravitationally bound to a surviving satellite at $t=0$, whereas empty histograms represent the entire accreted stellar mass (both disrupted and still gravitationally bound). The inner halo region in the simulations is the counterpart of the ``local halo'' typically accessible in observations. The figure shows that this region assembles early on. For example, $\sim 80\%$ of all inner halos (indicated by arrows in the second row panels) are already in place by look-back time $\sim 8-9$~Gyr.  The bulk ($\sim 80\%$) of the intermediate and outer regions form roughly at the same time as that of the inner one, however these regions have a slightly larger stellar fraction added at later times (10~Gyr $< t_{accr} <$ 5~Gyr) than the inner one.  The material still bound today (i.e. residing in surviving satellites) has been accreted more recently, generally within the past $5 - 6$~Gyr.  None of the surviving satellites are located in the inner $R<20$~kpc of the halo, where tidal forces are strongest. This result is in agreement with the location of present day Local Group dwarfs, which tend to reside at large distances from the center of the potential well \citep{mateo98} (the single exception is the Sagittarius galaxy, which is in the process of being disrupted).

Both the time of accretion and orbital parameters are expected to determine the chances for survival for infalling substructure. Figure \ref{fig:orbits} shows the accretion times, $t_{accr}$, and orbital circularities, $\epsilon$,  for baryonic satellites of all four halos, superimposed.  The orbital circularity $\epsilon$ is defined as the ratio of the angular momentum of the orbit of the satellite, $J$,  and of that of an equivalent circular orbit of the same energy, $J_{circ}$. Empty squares denote satellites fully disrupted  and filled hexagons satellites which are either intact or which still retain some bound material at present time. The figure suggests a consistent pattern for all halos: satellites accreted early, $t_{accr} >9$~Gyr, are fully disrupted by the tidal field, regardless of their orbital characteristics. Satellites accreted at intermediate look-back times, 5~Gyr $<t_{accr}<$ 9~Gyr, are more likely to survive if they are on or near circular orbits ($\epsilon \rightarrow 1$). All satellites accreted within the past 5~Gyr survive.  Applying our results to present day satellite galaxies of the Milky Way, we can infer that these are likely to have been accreted less than $8-9$~Gyr ago.

Our results suggest that present day satellites had more time available for sustaining star formation and hence further their chemical enrichment compared with local halo progenitors.  Hence  difference in chemical abundances between infalling substructure might be expected.

\subsection{Mass Accretion Histories}

We further investigate the relative contribution of different satellites to the stellar halo. Is the halo made up mainly by a few massive satellites or by a multitude of smaller ones? Figure \ref{fig:fhalo_mstar} shows the fraction of the stellar halo $f_{*}$  contributed by each disrupting satellite of stellar mass $M_*$, either binned in mass intervals $\Delta {\rm log}(M_{*}/M_{\odot}) =0.5$ (histograms) or by individual contributions (symbols). Solid lines and filled squares correspond to the entire stellar halo, and dashed lines and empty circles correspond to the inner halo. This figure shows that, by stellar mass, a few massive satellites contribute most to both the inner and the total halo.  One or a few more satellites in the range  $M_{*} \sim 10^{8} - 10^{10} \, M_{\odot}$ can make up $50\% - 80\%$ of the stellar halo. These results are in agreement with results of previous studies which found that most of the dark matter halo is made up of a few massive, $M \sim 1 - 10 \times 10^{10} \, M_{\odot}$, (dark matter) sub-halos \citep{zentner03,helmi03}, or that the properties of the present day population of globular clusters are consistent with most of these originating in a few massive halo progenitors \citep{cote00}.

\subsection{Chemical Abundances in the Stellar Halo and Surviving Satellites}
 
The link between the mass accretion histories and the chemical abundances of satellites for the four simulated halos is analyzed in Figure~\ref{fig:mass_metal}. The top panels for each halo show the distribution of average \FeH values (averages are weighted by the contribution of different stellar populations) versus stellar mass of satellites, $M_{*}$.  The different symbols separate satellites in terms of their accretion time: blue squares denote satellites accreted at lookback time $t_{accr} > 9$~Gyr, green triangles those with 5~Gyr $<t_{accr}<$ 9~Gyr, and magenta pentagons are used for satellites accreted less than 5~Gyr ago. The dashed lines in each top panels correspond to the stellar mass - metallicity fit derived by \citet{dekel03} based on Local Group dwarf data, confirming the agreement. 

This figure shows that there are many satellite galaxies accreted at early times with a wide range of \FeH, including a few massive ($M_{*} > 10^{9} M_{\odot}$) ones. Satellites accreted later have intermediate or large masses and generally larger average \FeH values. This is to be expected since both $M_*$ and \FeH increase monotonically with star formation. The middle panels repeat the stellar mass - \FeH data, this time emphasizing whether satellites are fully disrupted (triangles) or still retain some bound material at $t=0$ (hexagons). This shows that surviving satellites generally have larger average \FeH values, as might be expected from their late time of accretion. 

The bottom panels show the average \alphaFe values (also weighted by the contribution of different stellar populations) versus $M_{*}$ for all satellites in a given halo. Here $\alpha$ is taken as the average [(Mg + O)/2Fe] (this definition will be used thereafter in the paper). The figure shows that satellites which still have some bound material at present time also tend to have lower \alphaFe than the disrupted satellites. 

We also verify whether the resulting ages of the stellar populations in our simulated halo agree with observations. The top panel in Figure \ref{fig:age_metallicity} shows the age - metallicity relation for all disrupted satellites accreted onto one of the simulated halos (Halo H1), plotted as weighted average values. The result is in general agreement with the observed age - metallicity relation in the Milky  Way halo \citep{buser00}.  In the bottom panel we plot the corresponding age distribution of halo stars in the H1 simulation, shown for different lookback time intervals: $t_{accr} > 9$~Gyr, 5~Gyr$< t_{accr}<9$~Gyr, and $t_{accr}<5$~Gyr, respectively. This figure shows that only those satellites accreted recently (within the past $\sim$5~Gyr) contain significant young stellar populations, i.e. with ages less than $\simeq 5-7$~Gyr.
 
\subsection{Abundance Distribution Functions}

The metallicity distribution function (MDF) is frequently used to infer the chemical evolution of galaxies and, from there, their formation history. In the following, we analyze both the \FeH and \alphaFe distribution functions, which from now on will be broadly referred to as  ``abundance distribution functions'' (ADF).

Figure~\ref{fig:mdf1_halo_sats} shows a comparison between the \FeH distribution function of the surviving satellites (dashed lines) and that of the inner halos (solid lines) in our set of four simulations. 
The MDF peaks of the inner stellar halos range from $ \simeq -1$ (halos H1, H3 and H4) to $\simeq -0.6$ (halo H2). The \FeH distributions of satellite stars are typically narrower than the corresponding distributions of the inner halos and they too, show a range in their peaks: from $\simeq -1.4$ (halo H2) to $-0.4$ (halo H3).

\begin{figure}
\begin{center}
\epsfxsize=18cm\epsfysize=18cm\epsfbox{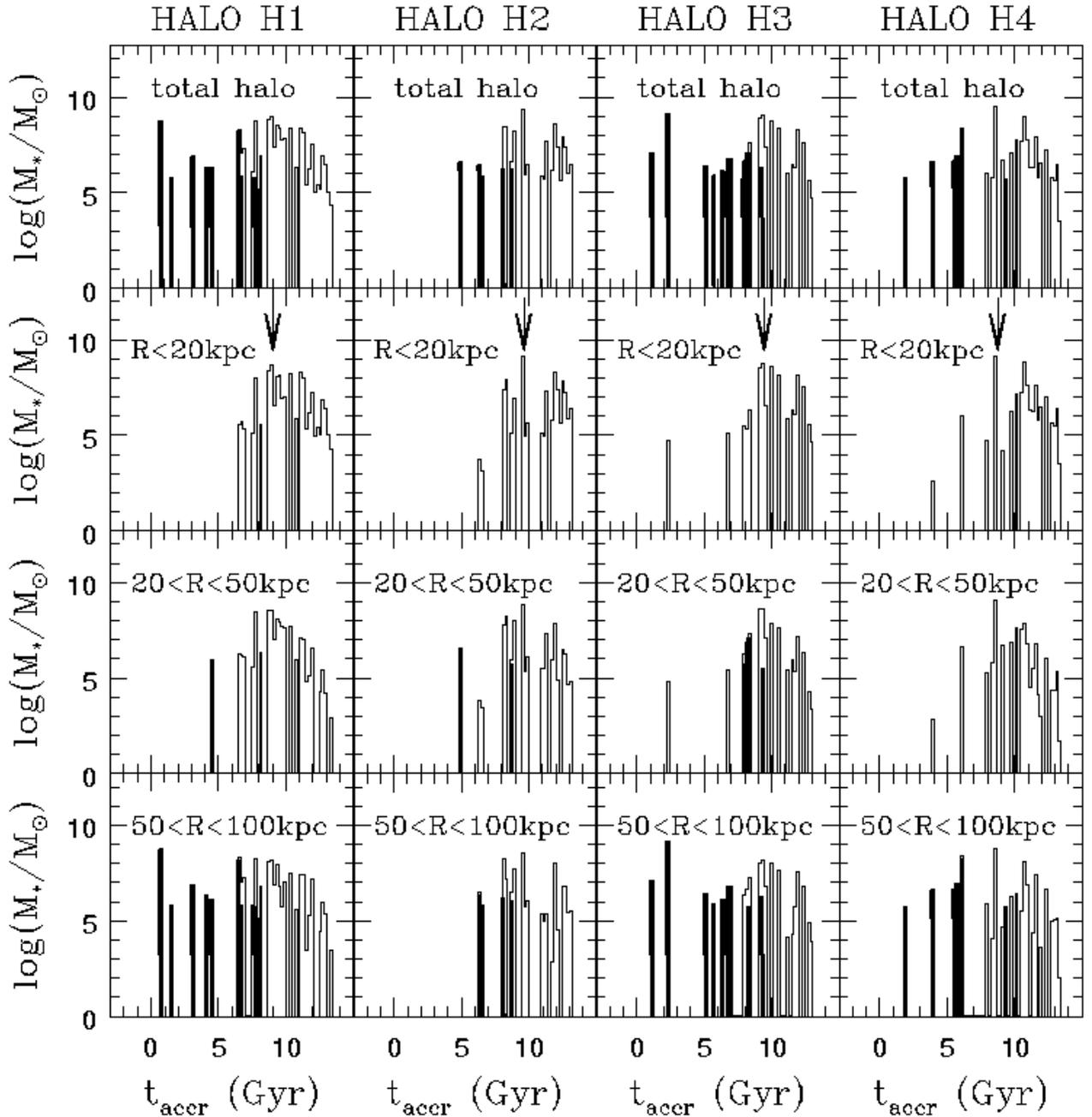}
\caption{\label{fig:mass_accr}{Mass accretion histories of the four simulated halos. The histograms represent the accreted mass which is deposited in the entire halo (first row); in the inner halo, $R<20$ kpc (second row); in the intermediate halo, $20$ kpc $<R<$ 50 kpc (third row); and in the outer halo, $50$ kpc $<R<$ 100 kpc (fourth row), respectively. Here $R$ is the distance relative to the position of the Sun, $R_{\odot} =8.5$~kpc. The empty histograms represent the entire mass (both bound and unbound) accreted in the corresponding distance interval and the filled histograms represent the material which is still gravitationally bound at t=0. The arrows in the second row panels correspond to the times when 80$\%$ of the inner halo has formed in each run.
}}
\end{center}
\end{figure}

\begin{figure}
\begin{center}
\epsfxsize=14cm\epsfysize=14cm\epsfbox{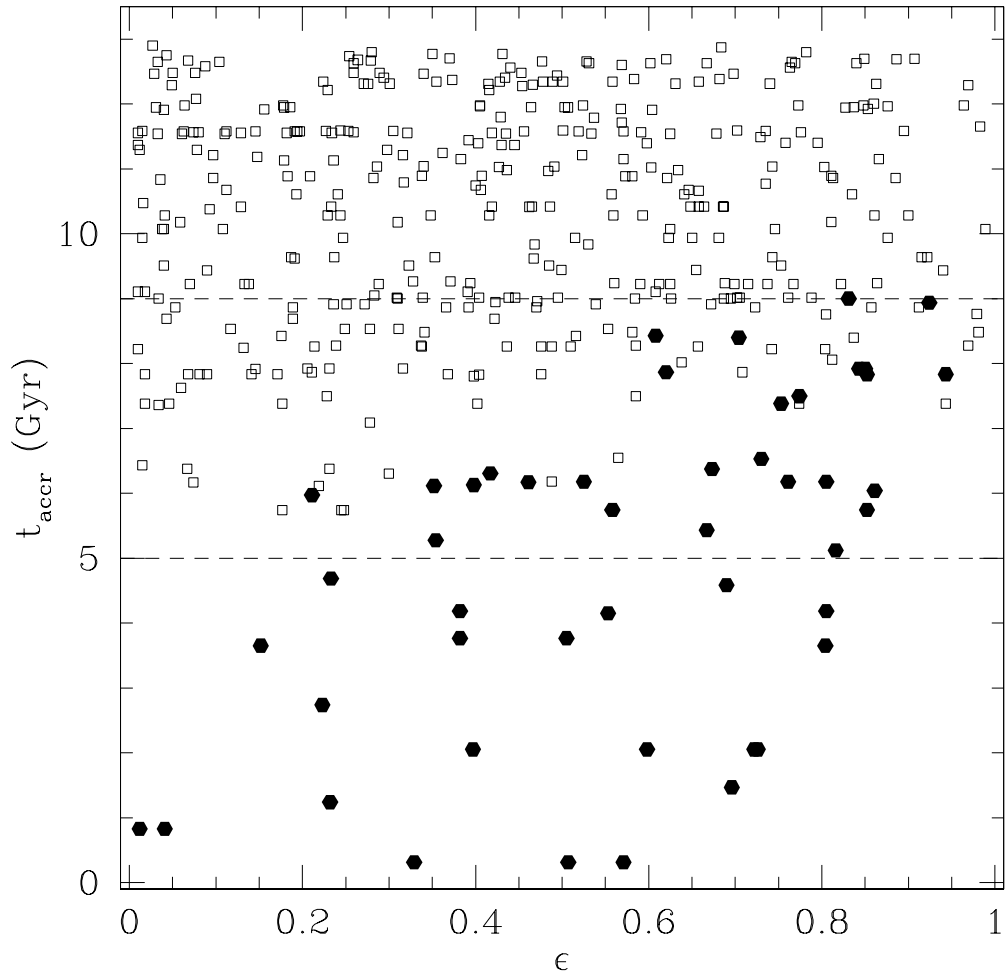}
\caption{\label{fig:orbits}{The time of accretion, $t_{accr}$, versus orbital circularity, $\epsilon$, are plotted superimposed for all satellites of the four simulated halos. Satellites totally disrupted at $t=0$ are plotted with empty squares and those that still have some bound mass at the same time are in filled hexagons.  Satellites tend to have more chances to survive if they are accreted late (lookback time $t_{accr} \le 9$~Gyr) and are on relatively circular orbits (circularity $\epsilon \rightarrow 1$). 
}}
\end{center}
\end{figure}

\begin{figure}
\begin{center}
\begin{tabular}{c}
\epsfxsize=8cm\epsfysize=8cm\epsfbox{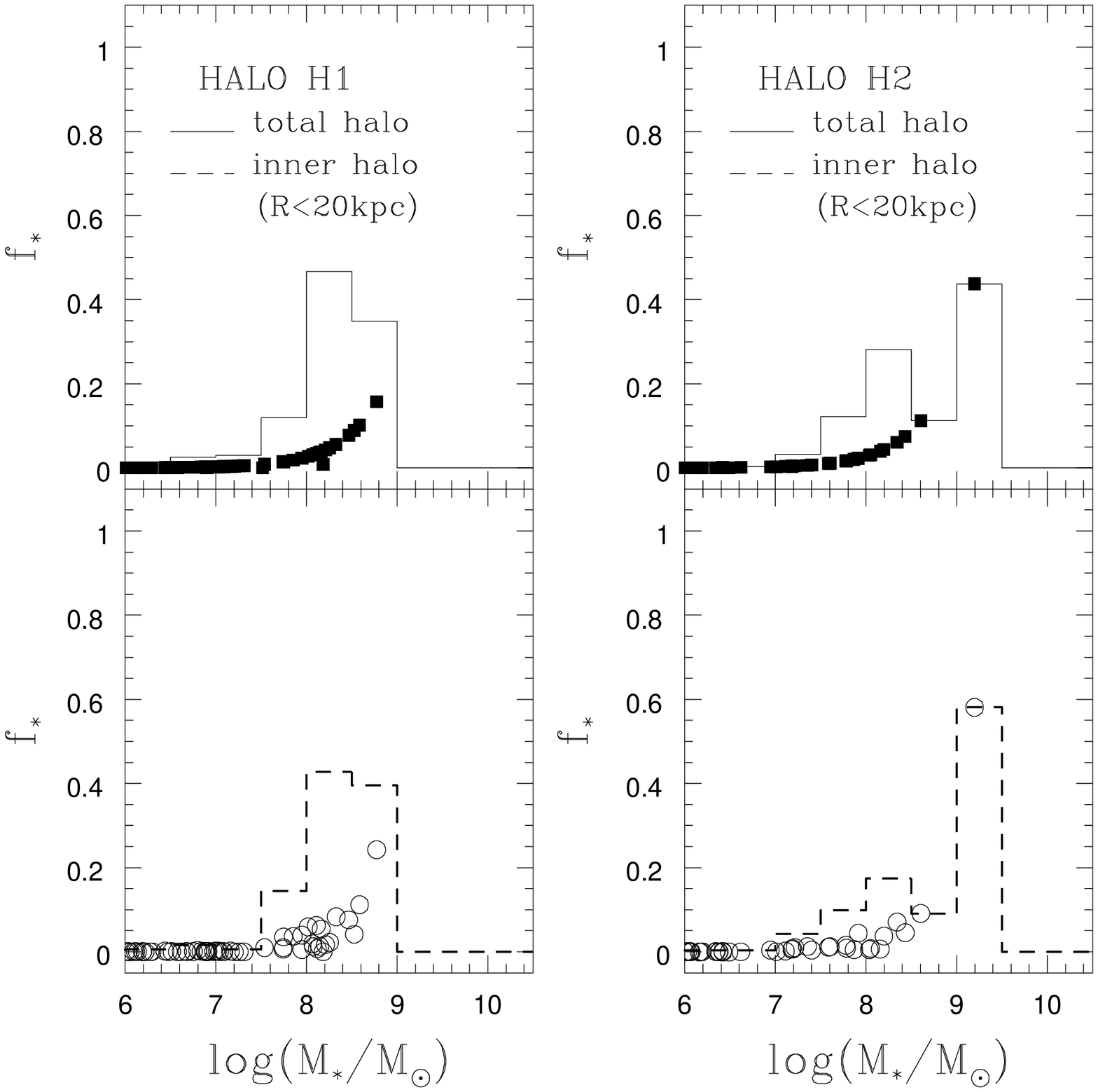}
\\
\epsfxsize=8cm\epsfysize=8cm\epsfbox{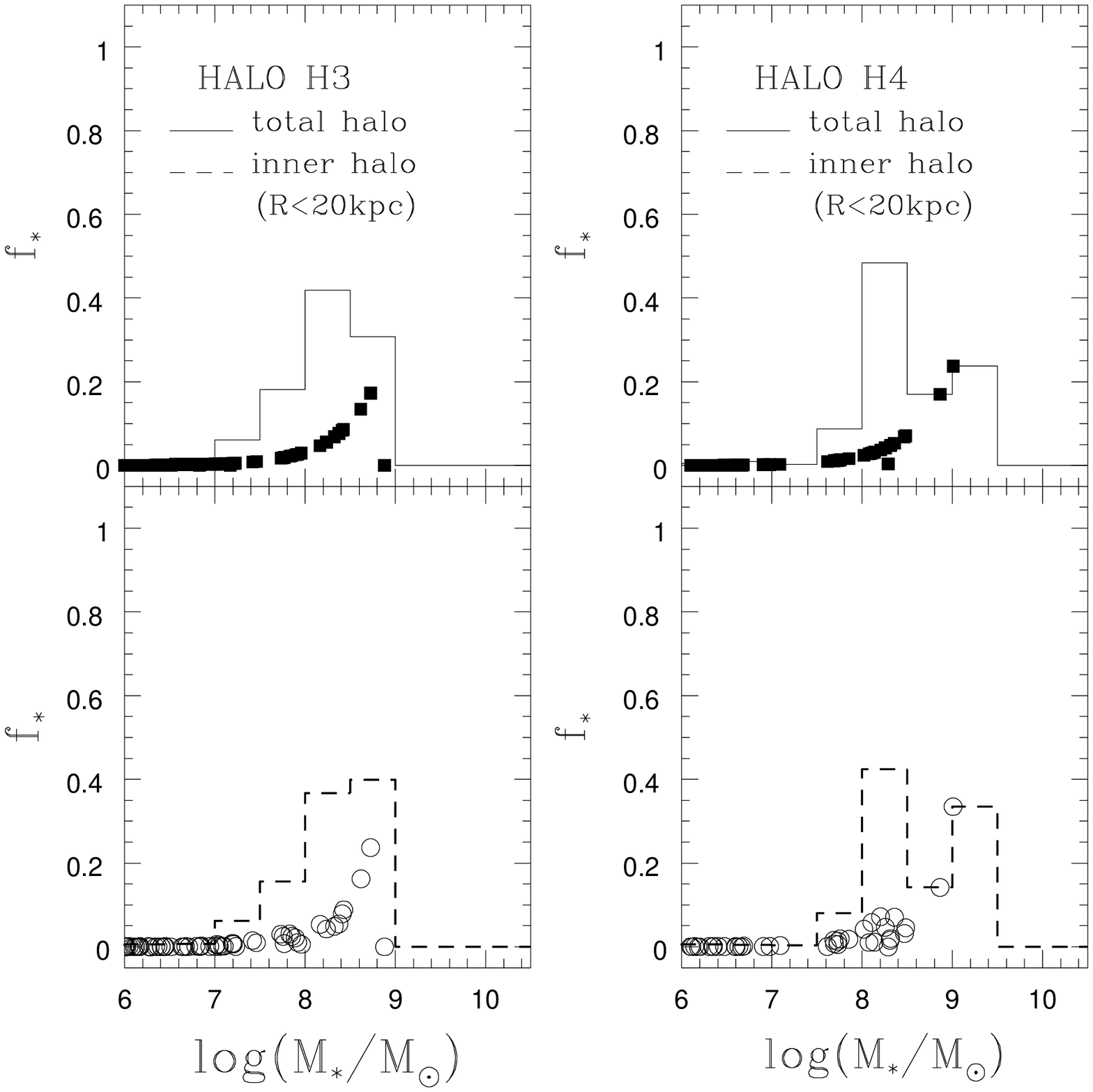}
\end{tabular}
\caption{\label{fig:fhalo_mstar}{Stellar fraction $f_{*}$ of the halo versus the stellar mass $M_{*}$ of the contributing satellites. Solid lines and filled squares correspond to the satellites contributing to the total halo; and the dashed lines and empty circles to those which contribute to the inner halo.
}}
\end{center}
\end{figure}

\begin{figure}
\begin{center}
\begin{tabular}{cc}
\epsfxsize=8.cm\epsfysize=8.cm\epsfbox{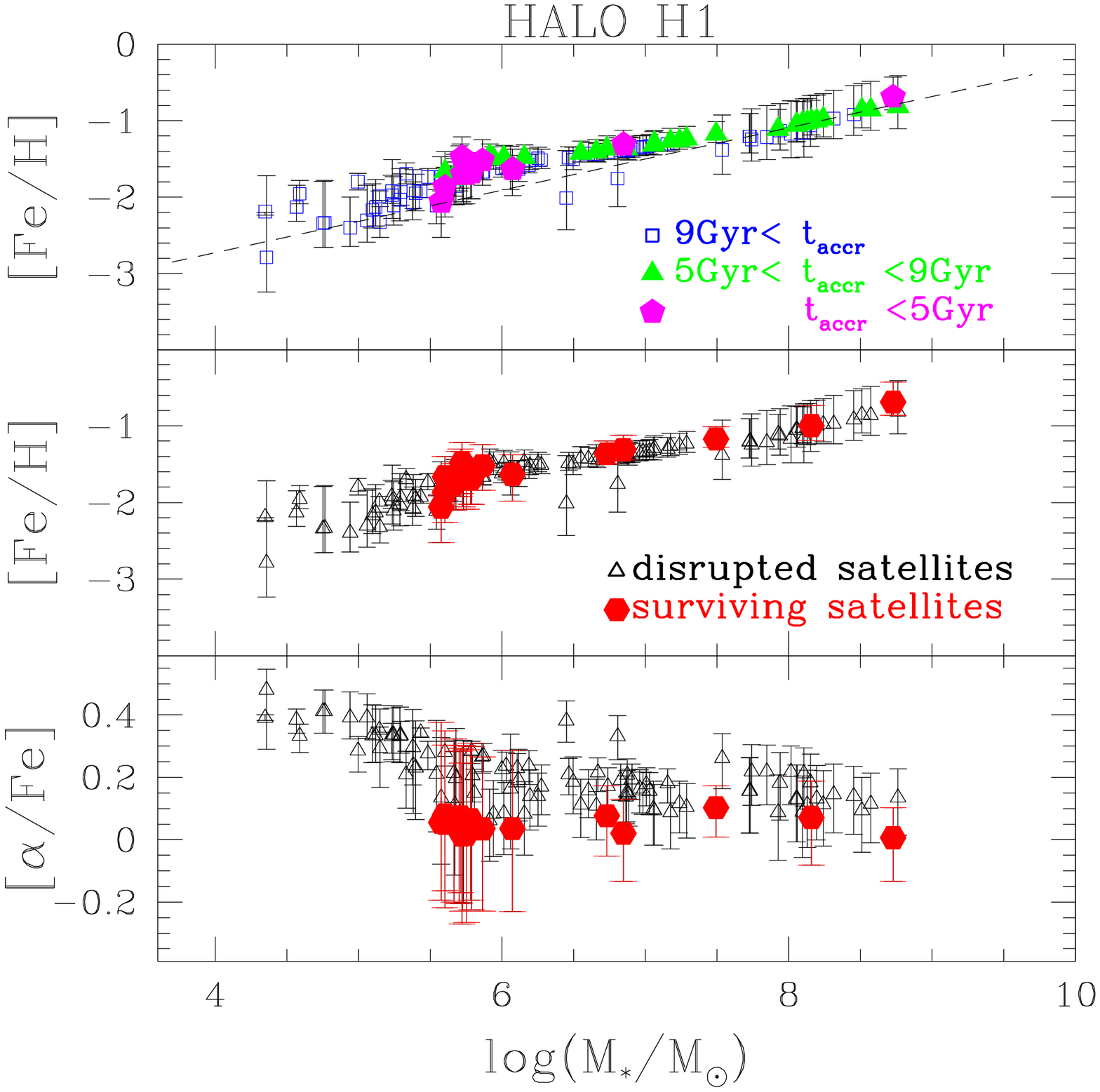} &
\epsfxsize=8.cm\epsfysize=8.cm\epsfbox{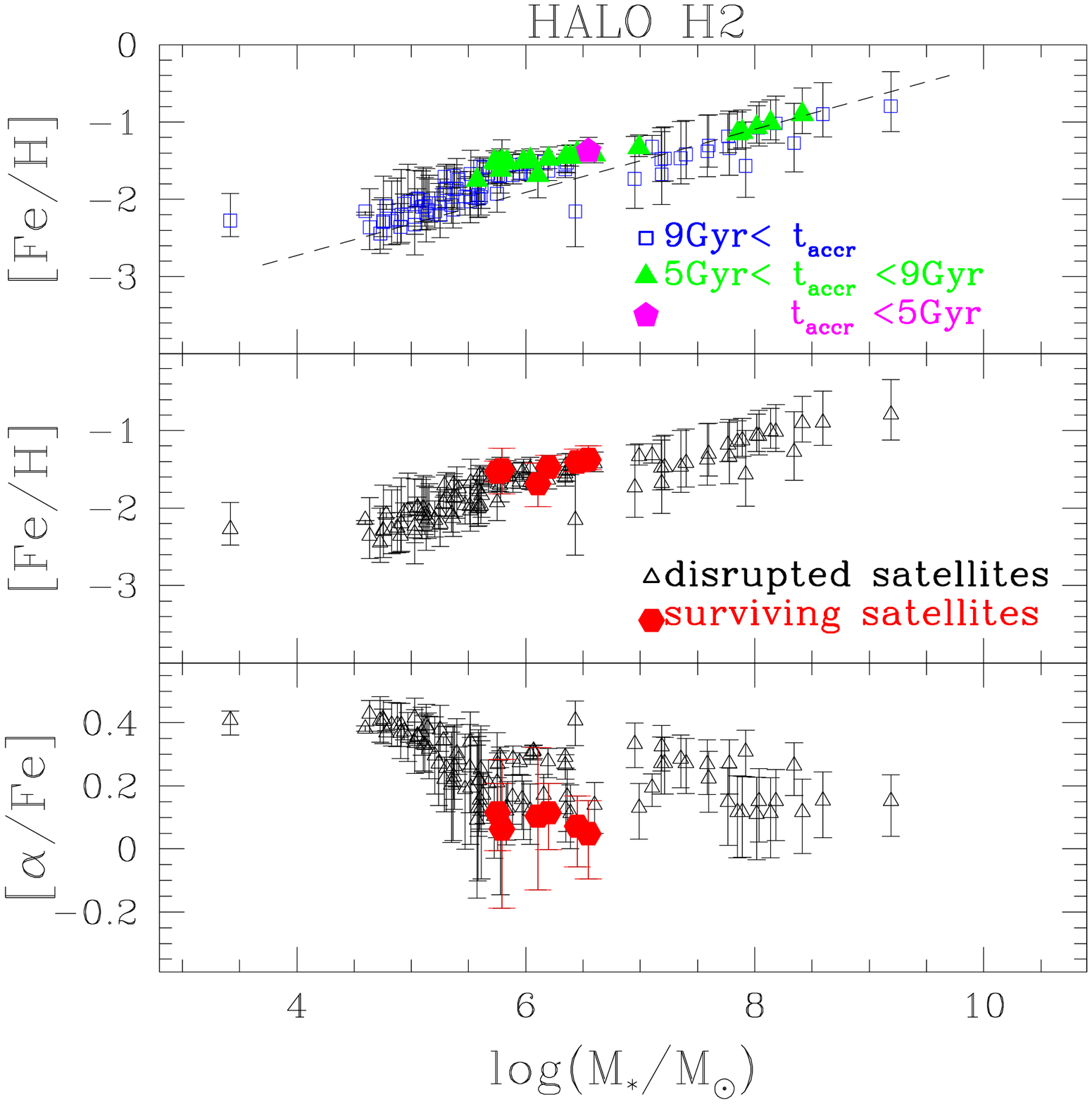} \\
\epsfxsize=8.cm\epsfysize=8.cm\epsfbox{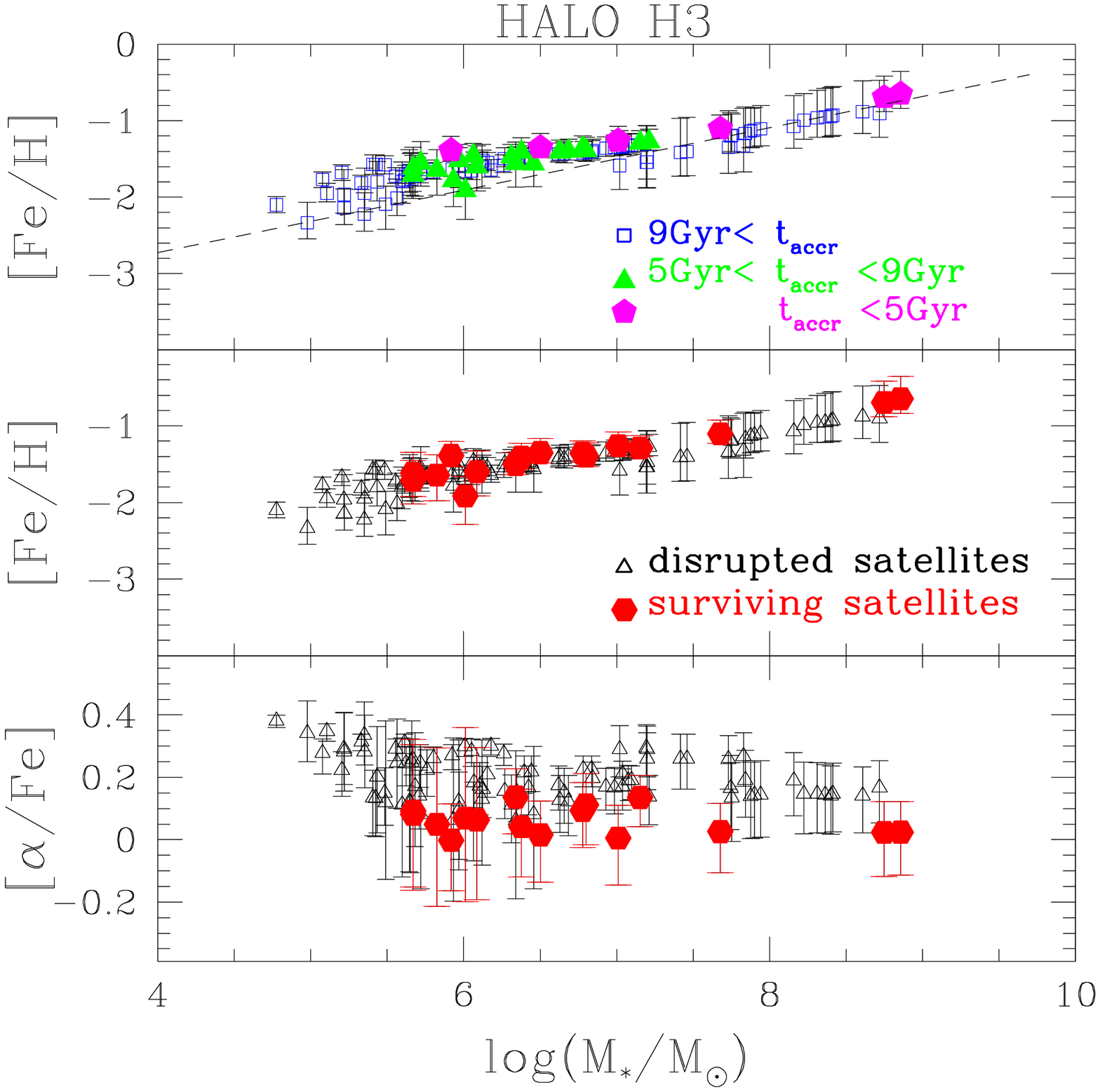} & 
\epsfxsize=8.cm\epsfysize=8.cm\epsfbox{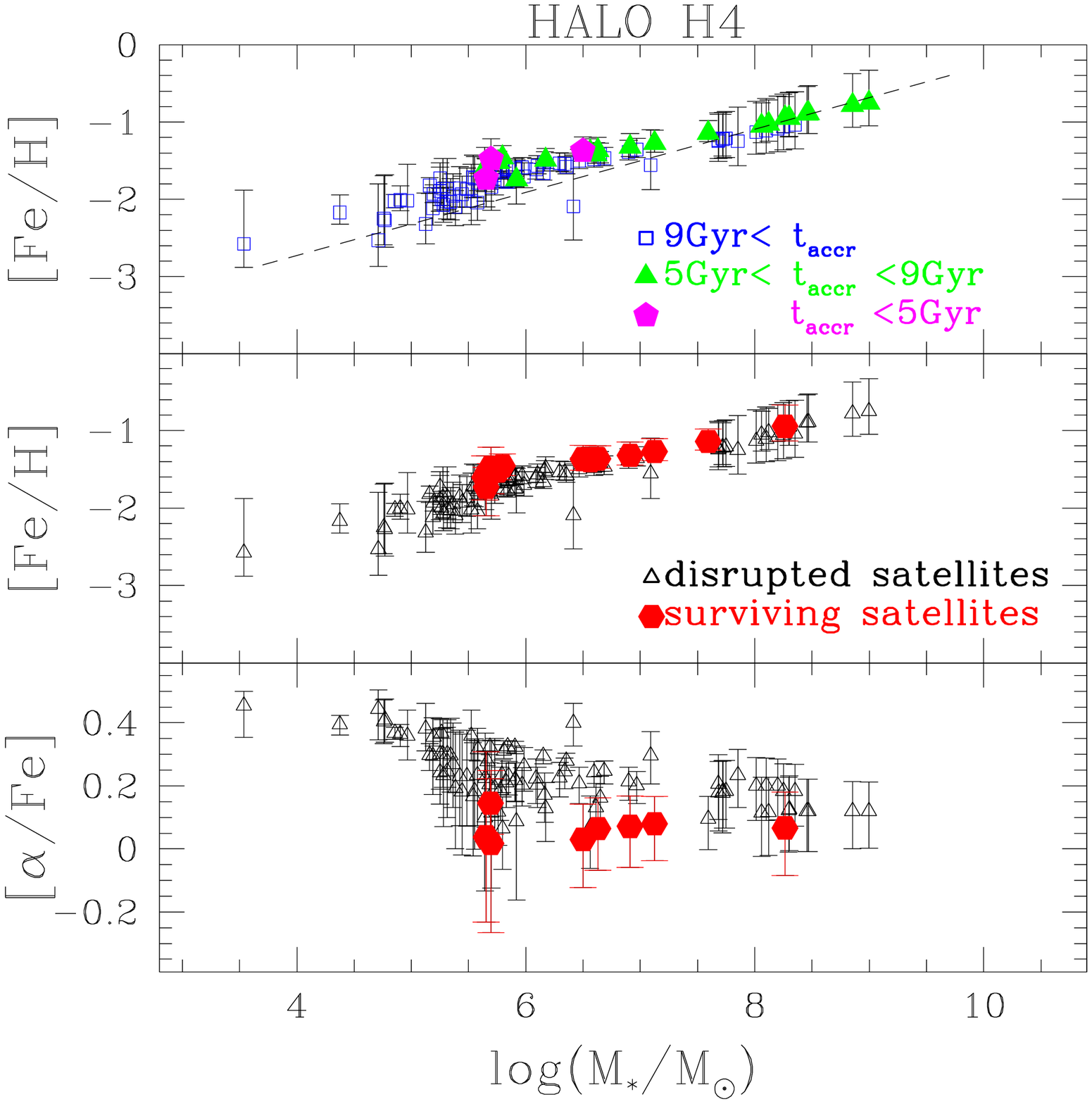}
\end{tabular}
\caption{\label{fig:mass_metal}{Chemical abundance patterns for the satellite distribution in the simulated halos. Each panel corresponds to a simulated halo and contains three sub-panels. The first sub-panel from the top illustrates average \FeH  values versus stellar mass, $M_{*}$, color coded in terms of their time of accretion: blue squares for satellites accreted at lookback time $t_{accr} > 9$~Gyr, green triangles for satellites with 5~Gyr $<t_{accr}<$ 9~Gyr, and magenta pentagons for satellites accreted less than 5~Gyr ago.  The dashed line in each top subpanel corresponds to the \citep{dekel03} relation. The second sub-panel contains the same parameters, this time emphasizing the  satellite material still gravitationally bound at $t=0$ (red hexagons) versus satellites which are totally disrupted (black triangles). The third subpanel shows average \alphaFe values versus $M_{*}$ for the satellite material still bound at $t=0$ (red hexagons) and for the satellites which are totally disrupted (black triangles).
}}
\end{center}
\end{figure}

\begin{figure}
\begin{center}
\begin{tabular}{c}
\epsfxsize=9cm\epsfysize=9cm\epsfbox{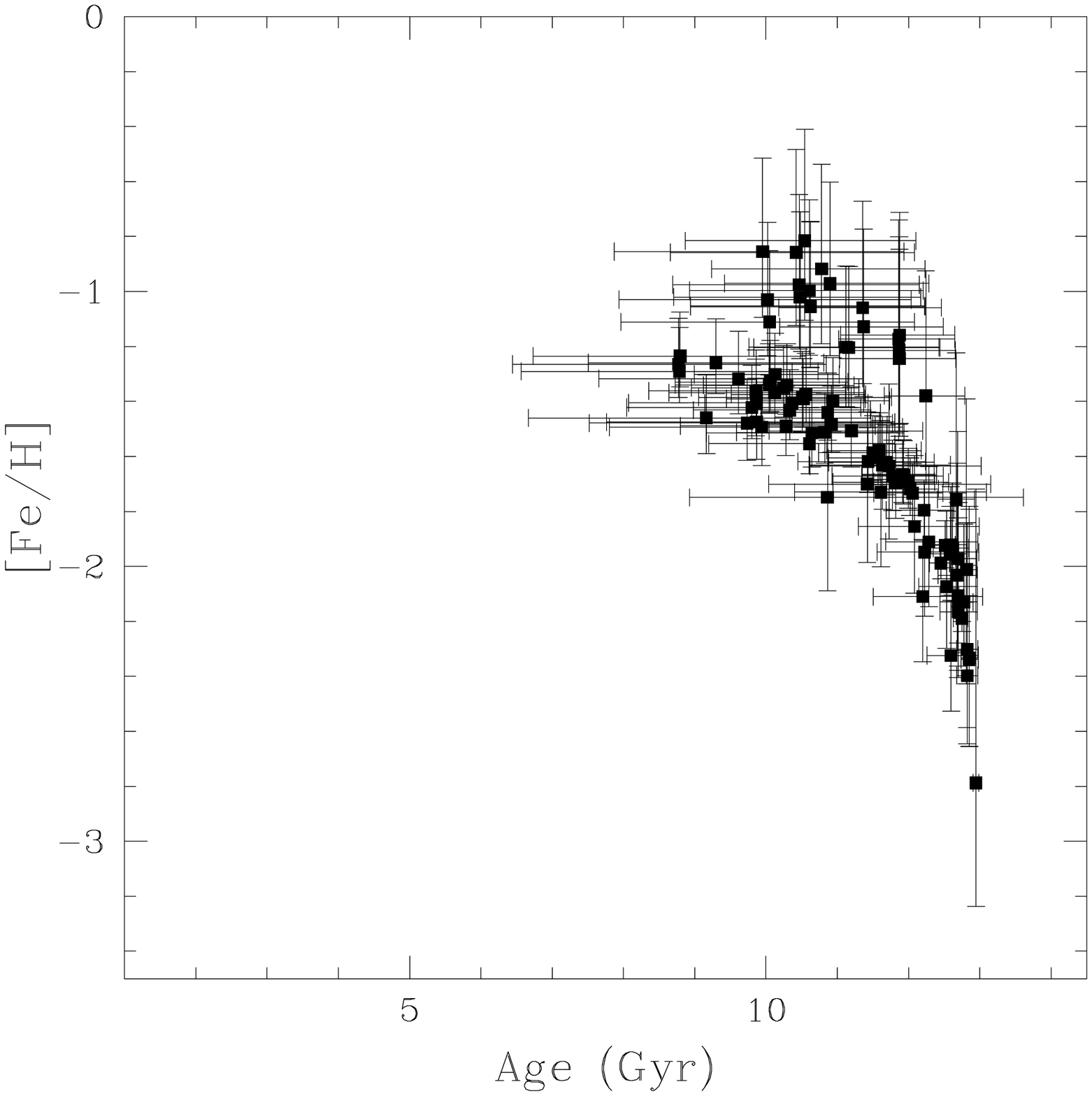}\\
\epsfxsize=9cm\epsfysize=9cm\epsfbox{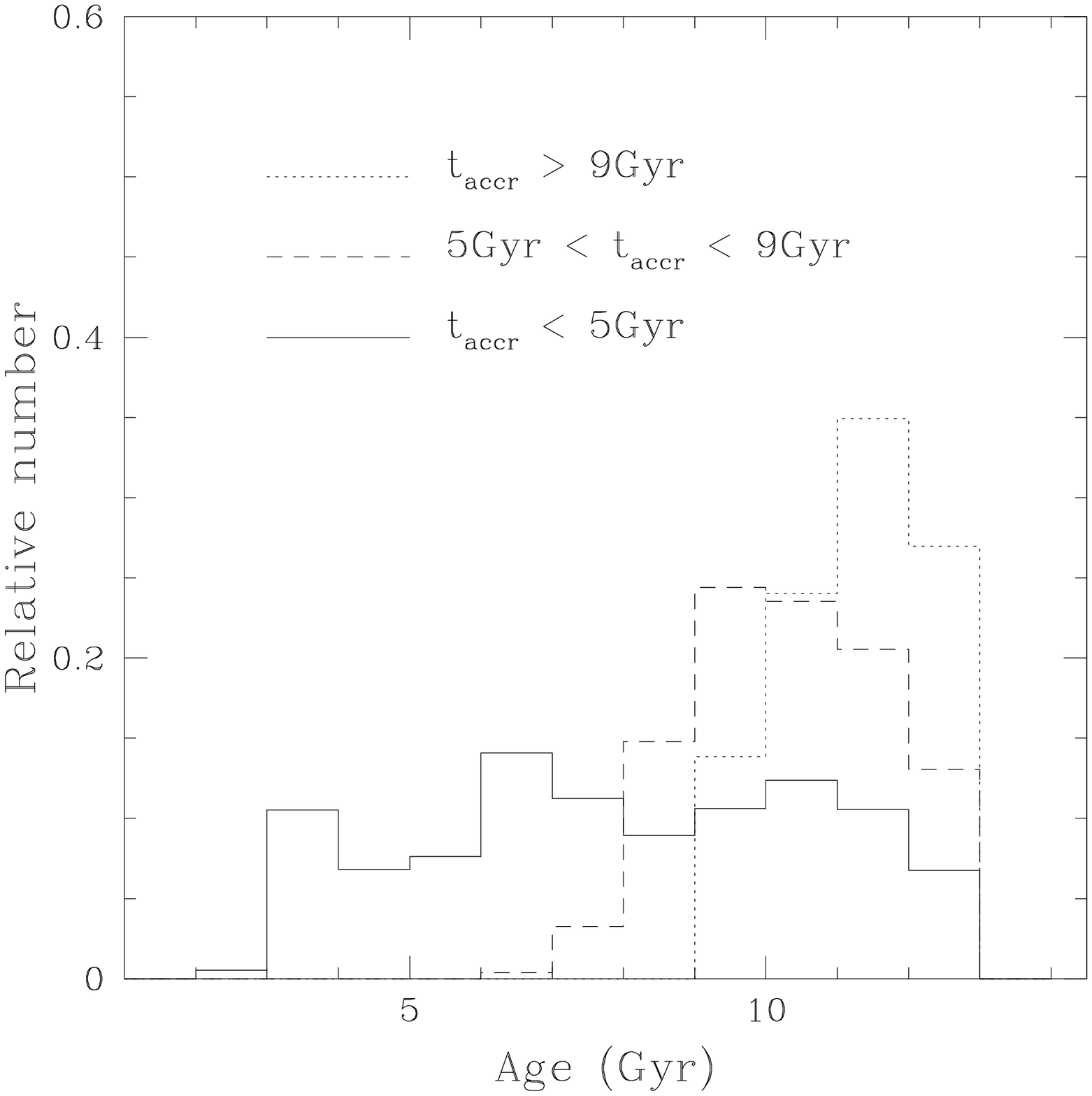}
\end{tabular}
\caption{\label{fig:age_metallicity}{{\it(Top panel)} Age - metallicity relation for one of the simulated halos (Halo H1). The data points represent weighted averages for all satellites with errorbars spanning from $10\%$ to $90\%$ of these values. {\it (Bottom panel)} The age distribution of halo stars accreted more than 9 Gyr ago, between 5 and 9 Gyr ago and within the last 5 Gyr. The histograms are normalized to each distribution.}}
\end{center}
\end{figure}

\begin{figure}
\begin{center}
\epsfxsize=16cm\epsfysize=16cm\epsfbox{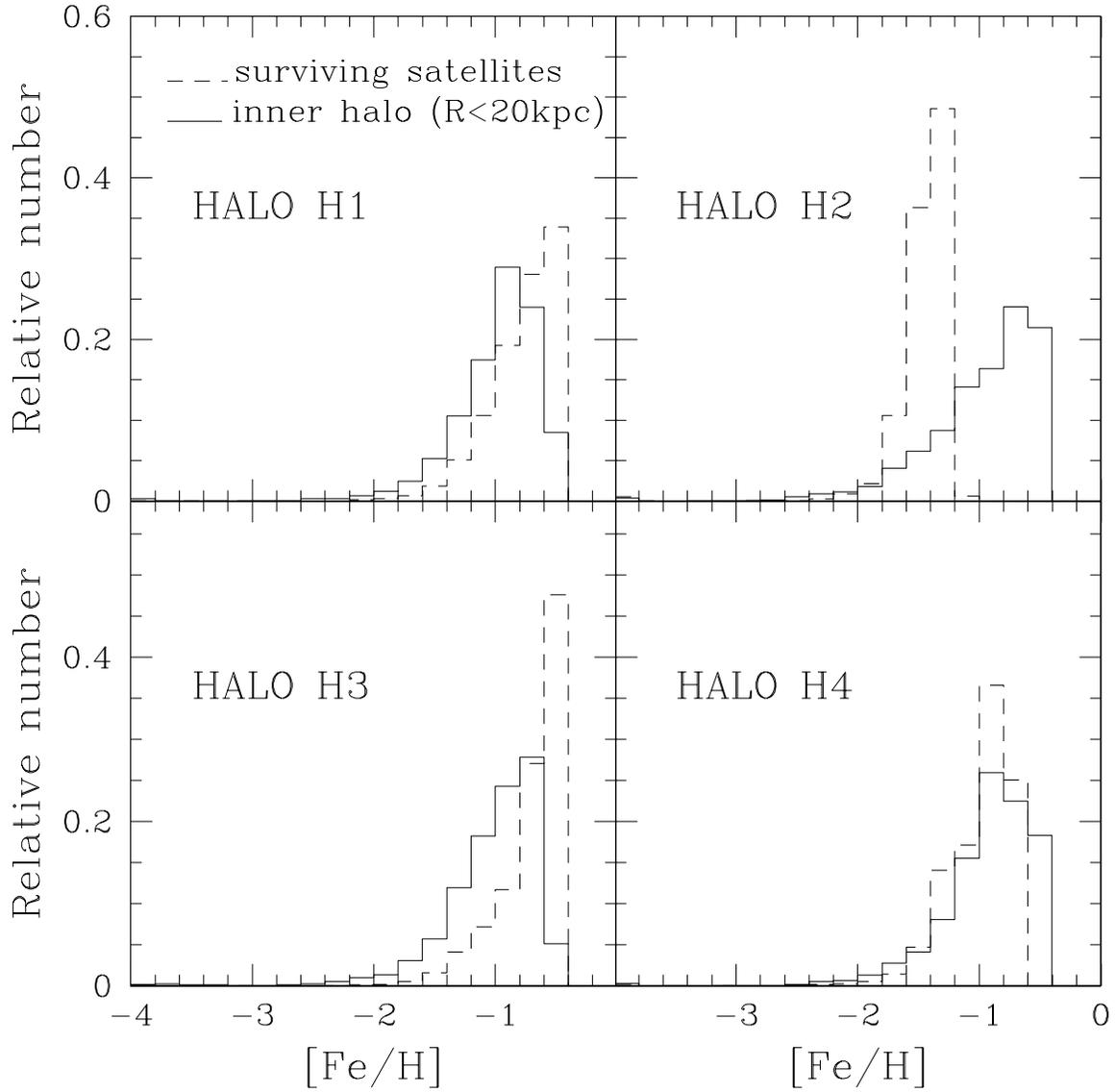}
\caption{\label{fig:mdf1_halo_sats}{The \FeH distribution functions of stars in the inner halo (solid line) and of those in surviving satellites (dashed lines) for the four simulations.
}}
\end{center}
\end{figure}

The differences in the location of the MDF peaks in the four simulations arise from the different halo mass accretion histories. For example, most of the surviving satellites in the H2 run have been accreted earlier than the corresponding surviving satellites in other runs (see Figure \ref{fig:mass_accr}), and have low masses, hence relatively low \FeH values (see Figure \ref{fig:mass_metal}). Therefore the surviving satellites \FeH peak is shifted toward the more metal poor end. On the other hand, most of the contribution to the inner H2 halo MDF is made by as single massive satellite, $M_{*} \simeq 10^{9} \, M_{\odot}$ (see Figure \ref{fig:fhalo_mstar}), whereas a significant fraction to the H1 and H3 inner halos is added by numerous smaller satellites. This implies a shift in the \FeH distribution peak towards the more metal poor values in halos H1 and H3 than in H2. The H4 run is the only one with a similar \FeH distributions in the inner halo and in surviving satellites, which is the result of a balanced contribution to the halo stellar fraction from both low and high mass satellites (see Figure \ref{fig:fhalo_mstar}) and from an average mass range of the surviving satellites, $M_{*} \simeq 10^{8} - 10^{9} \, M_{\odot}$ (see Figure \ref{fig:mass_accr}).

Figure~\ref{fig:mdf2_halo_sats} shows a similar histogram for the \alphaFe data. As for \FeH, the \alphaFe distribution of satellite stars is narrower than the corresponding distribution of the inner halo, and generally shows a systematic offset. The spread within the \alphaFe distribution function seems to be consistently  smaller than the spread in the \FeH distribution function, for both surviving satellites and inner halo.

\section{Discussion}

\subsection{Comparing the Milky Way and simulated chemical abundance patterns.}

 How do our Milky Way-type halos fare in comparison with the Milky Way data? Figure \ref{fig:mgfe_feH_venn} shows the \MgFe versus \FeH from the compilation data\footnote{Note that the \citet{venn04} data do not contain \OFe, therefore we limit our comparison only to \MgFe.} of \citep[][see the references therein for information about the original data]{venn04}. This sample contains stars from both halo and dwarf galaxies and a wide range of chemical abundance information. We select from this compilation all satellite stars (in total of  36)  and  all stars with halo membership probabilities greater than $50\%$, which results in a total of  259 halo stars (compared with 235 halo stars, had we selected all stars with halo membership probability of $100\%$). 

In  Figure~\ref{fig:alpha_halo_sats} we plot the  \MgFe versus \FeH for the simulated local halos and surviving satellites. For the local halo, we use the unbound stellar material binned in \FeH =0.1~dex intervals. The values corresponding to the surviving satellites are weighted averages only over the bound material to these systems. Solid lines represent errorbars of 10$\%$ and 90$\%$ of weighted average \MgFe values, respectively. Dotted lines highlight the absolute extent in \MgFe values for the inner halo. Figure \ref{fig:alpha_halo_sats} shows again that stars in surviving satellites have consistently lower \MgFe values than stars in the halo. Although there is some overlap in the \MgFe values, the great majority of halo stars have \MgFe values $0.1 - 0.4$~dex higher than that of present day satellites\footnote{The chemical abundance results shown throughout the paper are based on a modeling with the Kroupa initial mass function (IMF). However, we find that the choice of the IMF does not change significantly the results. Using a Salpeter IMF gives essentially the same abundance patterns and distribution functions. Note however that to obtain the same normalization discussed in \S 2.2, the model with the Salpeter IMF requires less feedback or shorter star formation rate (see \citet{robertson05} for details).}.

 Our results  reproduce  the   trend of Galactic  chemical   abundance patterns found  in observations and the  overall offset in \MgFe (and  hence \alphaFe) ratios  between  satellites disrupted  and  surviving
 dSphs.  As also discussed by \citet{robertson05}, the key factors that influence this offset are the time of accretion and the mass of the satellite contributing to the halo.  The \alphaFe  ratio   has been  often referred   to as  a  ``cosmic clock''  for  galaxy formation.  Based  on  the time delay of about 1~Gyr between  the onset of supernovae  Type II (which  produce both Fe and $\alpha$ elements) and Type Ia  (which produce Fe alone), a break in  \alphaFe ratios  of halo stars   is expected to  occur at  intermediate   metallicities (around \FeH  $\simeq -1.0$) \citep[see,  for example][]{mcwilliam97}.  The local halo is built up fast, within the first few Gyr of the lifetime of the  galaxy, from dwarf galaxies with high \alphaFe content (with a significant contribution from the few high mass systems accreted early on). Dwarf galaxies that survive until present time  were generally accreted later  (within  the last $5-6$~Gyr) and had time to be enriched in Fe, hence their lower \alphaFe.

A slight modification to the simple "cosmic clock" picture (which includes the effect of mass) allows us to explain the high \FeH/high \alphaFe component of the halo. In the framework of the hierarchical structure formation scenario, it is more appropriate to think of a multitude of ``clocks'' running at different rates and stopping at different times, associated with the individual star formation sites (i.e progenitors of the halo). Most of these clocks are stopped either on or just after accretion of the satellites. Since the bulk of the halo formed from {\it massive} satellites accreted early, these had very efficient star formation and rapid enrichment to high \FeH (i.e. the cosmic clock in this case runs fast, but stops early). In contrast, the dSph had inefficient star formation, shut off at much later times.

Lastly, this discrepancy in accretion times  provides a natural explanation for the lack of a counterpart in the halo to the intermediate age and young stellar populations seen  in surviving satellites: the halo was in place sufficiently long ago ($>8$~Gyr) that it does not contain a significant number of such stars.
 
Overall,  we have  shown that once the full dynamics and chemical evolution are taken into account, the $\Lambda$CDM  model can easily accommodate the chemical abundance patterns and stellar populations seen in the stellar halo and dSph.  
 
 \subsection{The Abundance Distribution Functions of the Milky Way}
 
\begin{figure}
\begin{center}
\epsfxsize=16cm\epsfysize=16cm\epsfbox{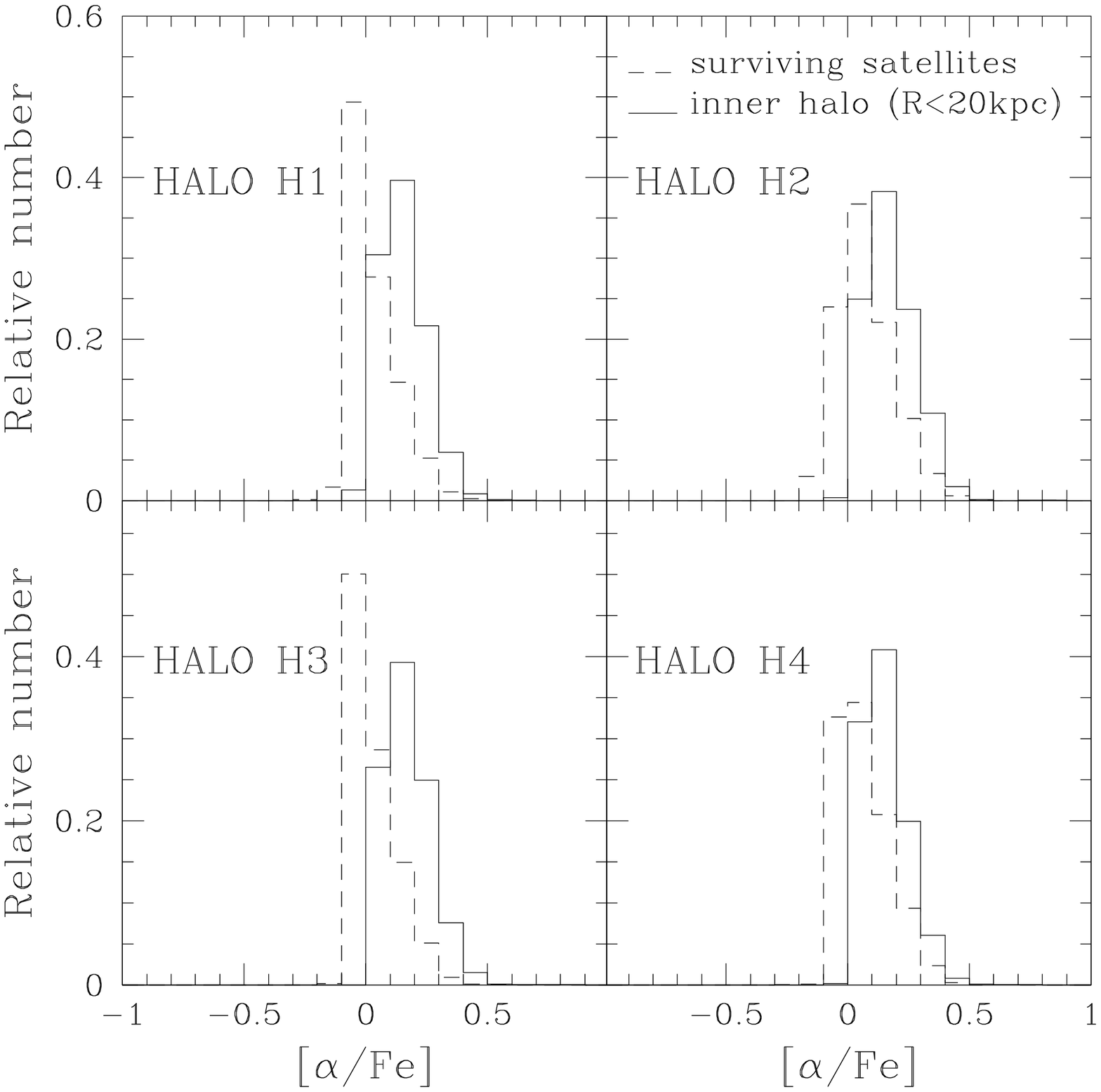}
\caption{\label{fig:mdf2_halo_sats}{The \alphaFe distribution functions of the stars in the inner halo (solid line) and those in surviving satellites (dashed line) for the four simulations.
}}
\end{center}
\end{figure}

\begin{figure}
\begin{center}
\epsfxsize=12cm\epsfysize=12cm\epsfbox{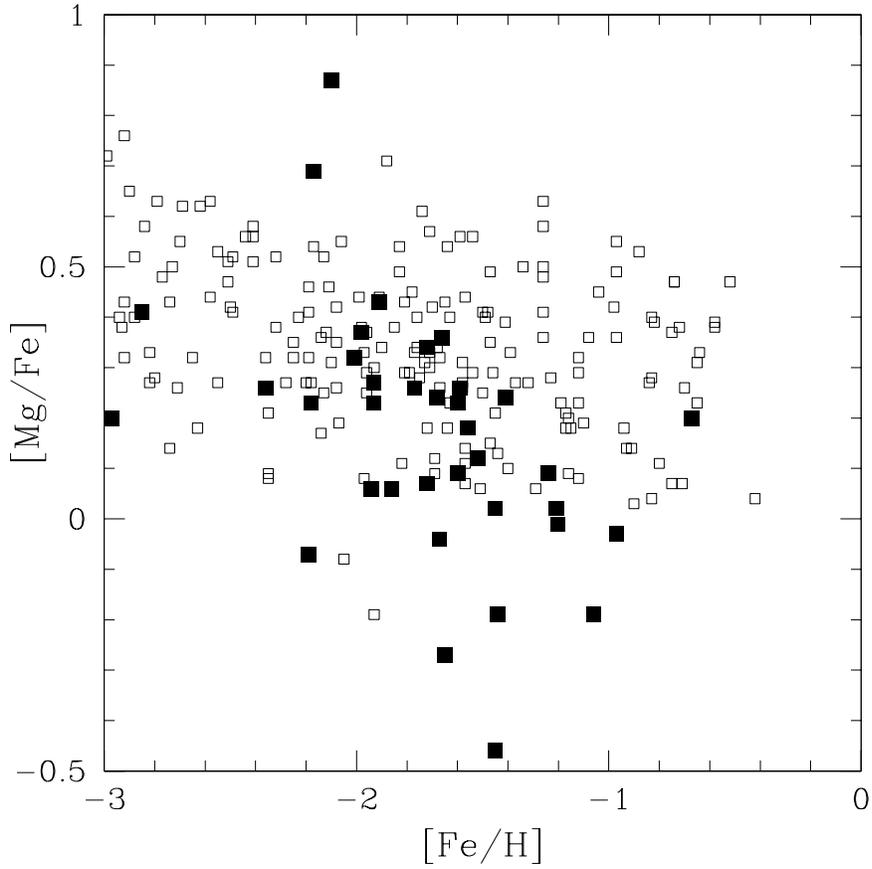}
\caption{\label{fig:mgfe_feH_venn}{The \MgFe versus \FeH for the \citet{venn04} compilation data. There are in total 36 satellite stars. For the halo, only the stars with halo membership probabilities greater than $50\%$ are considered.
}}
\end{center}
\end{figure}

\begin{figure}
\begin{center}
\epsfxsize=16cm\epsfysize=16cm\epsfbox{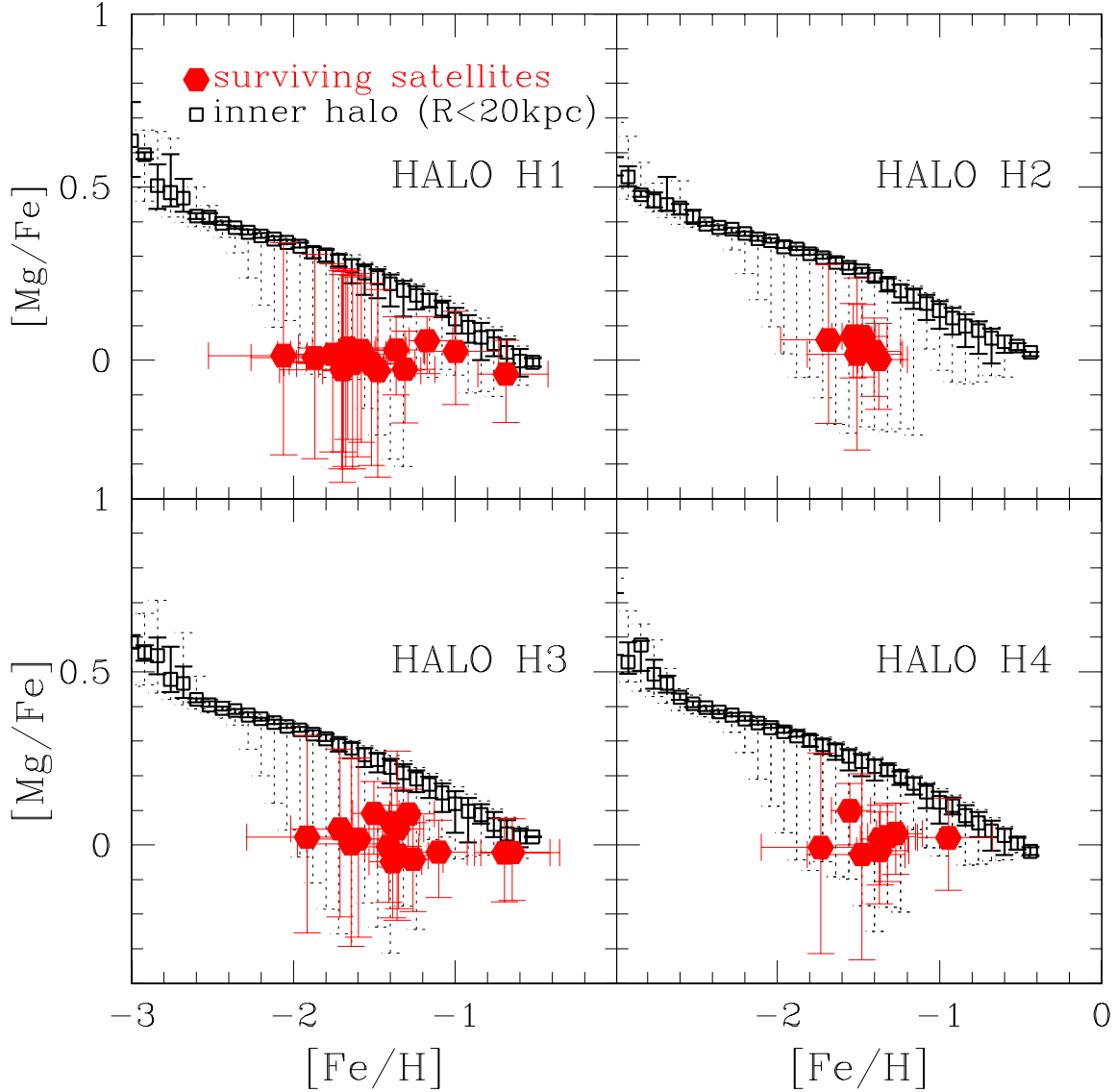}
\caption{\label{fig:alpha_halo_sats}{\MgFe versus \FeH for the sample of four halo simulations (values are weighted averages). Empty squares correspond to inner halo ($R<20$~kpc) values binned in $\Delta$\FeH$=0.2$~dex, and red hexagons correspond to values for surviving satellites. Solid lines represent errorbars delineating $10\%$ and $90\%$ from the absolute spread in \MgFe values, respectively.  Dotted lines highlight the absolute extent in \MgFe values for the inner halo. 
}}
\end{center}
\end{figure}

We note that the ADFs of our simulated data can only be cautiously compared with the available observations.  Our ADFs  (Figures \ref{fig:mdf1_halo_sats} and \ref{fig:mdf2_halo_sats}) are normalized functions over {\it all} stars in the local halo or surviving satellites, whereas the observational distributions are normalized functions only over a limited number of stars. For the stellar halo the observational challenge is to select a truly random sample of stars. For the satellites, the current high resolution spectral data is limited to a few stars each in some subset of the satellites, so it is impossible to construct a luminosity-weighted histogram.

Numerous MDFs have been derived for the Milky Way halo \citep{hartwick76,laird88,ryan91,carney96,chiba00}.  These studies have found that the peak of the halo MDF occurs around \FeH $=-1.5$.  In contrast, metallicity data for dwarf galaxies in the Local Group (both satellites and in the field) are still sparse \citep[see][and references therein]{grebel03,venn04}, but recent and upcoming observations are aiming to bridge that gap \citep{kaufer04,tolstoy04}.

\begin{figure}
\begin{center}
\epsfxsize=16cm\epsfysize=16cm\epsfbox{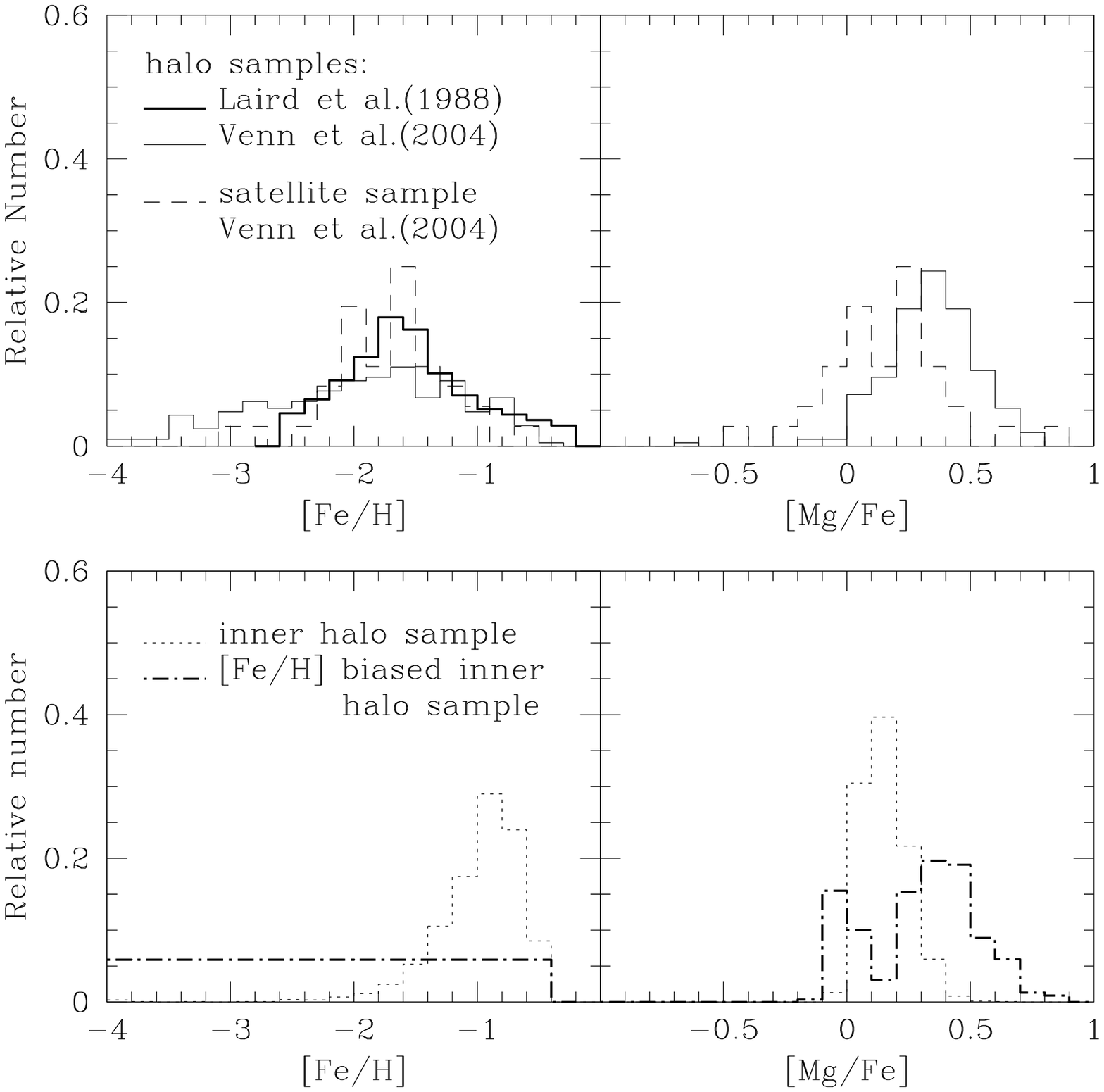}
\caption{\label{fig:mdf_obs_halo_sats_new}{{\it (Top panel)} \FeH distribution functions (top left) and [Mg/Fe] distribution functions (top right) based on a compilation of observational data. The thick solid line corresponds to the sample of halo stars of \citet{laird88} (note: this sample contains only \FeH data).  Thin solid lines  and dahsed lines correspond to the ADFs of halo and satellite stars from the compilation data of \citet{venn04} (see the references therein for information about the original data).  For the latter sample data, we consider only the stars whose halo membership probabilities are greater than $50\%$. {\it (Bottom panel)}  Normalized histograms showing the \FeH  (bottom left) and [Mg/Fe]  (bottom right) distribution functions of stars in the inner H1 halo. Dotted lines represent the entire collection of halo stars in the inner halo, whereas dot dashed lines correspond to a biased selection of equal number of stars per \FeH bin.}}
\end{center}
\end{figure}

The top left panel in Figure \ref{fig:mdf_obs_halo_sats_new} shows \FeH distributions based on a compilation of observational data, whereas the top right panel shows the \MgFe distributions. The thick solid line corresponds to the sample of halo stars of \citet{laird88} (note: this sample contains only \FeH data).  With thin dashed lines we are plotting the abundance distributions of satellite stars from the \citet{venn04} data. The available observational data seem to suggest that the two \FeH distributions (halo and surviving satellites) have their peaks at about the same location, around $-1.5$~dex. In the \citet{venn04} data, the satellite stars \FeH distribution seems to be narrower than that of the Milky Way halo, up to 1~dex. This is mainly because the \citet{venn04} satellite \FeH distribution seems to lack the very metal poor tail (\FeH $\sim -4$~dex) seen in the corresponding halo stars distribution. However, the \citet{laird88} halo data do not show the metal poor tail and this sample is, unlike the \citet{venn04} data, unbiased in metallicity. Moreover, it is unclear whether the lack of metal poor stars in Local Group dwarfs is real or just the result of incompleteness (at present there are only a few stars per satellite with measured chemical abundances). Stars in our model satellite galaxies also seem to have slightly lower \alphaFe (see Figures \ref{fig:alpha_halo_sats} and \ref{fig:mdf2_halo_sats}). This is in agreement with the observations (Figure \ref{fig:mdf_obs_halo_sats_new}; and Figure~2 in \citet{venn04}). 

As discussed above, the \citet{venn04} sample contains an intrinsic bias towards selecting more metal poor stars than in a random selection (the goal of their study was not to produce an MDF but to identify the \alphaFe $-$ \FeH trend, which necessitates a rather uniform sampling of the \FeH range). Therefore the \citet{venn04} data is not appropriate for retrieving the halo \FeH distribution function.  However, a bias in selecting \FeH does not necessarily translate into a bias in the \alphaFe.  The \alphaFe distribution function peaks are very narrow and unlikely to be missed by the \citet{venn04} study. We test this hypothesis by simulating a biased selection of halo stars of drawing from one of our simulated halos (halo H1) an equal number of stars in each \FeH bin and checking by what extent  this choice affects the  detection of the \alphaFe distribution function peak (note: this case is more extreme than the selection of \citet{venn04}).  Bottom panel in Figure \ref{fig:mdf_obs_halo_sats_new} shows the resulting normalized histograms representing the \FeH and [Mg/Fe] distribution functions of stars in the inner halo. Dotted lines represent the entire collection of inner halo stars and dot dashed lines correspond to the biased selection (in total, the stars selected this way amount to less than $1\%$ of the total stellar mass of the inner halo).  The  figure shows that, even in this extreme case, the [Mg/Fe] distribution function is remarkably close to the original one in retrieving the overdensity of stars with \MgFe $\sim 0 - 0.5$ values. This result give us confidence that the ADF peak revealed in the [Mg/Fe] data of \citet{venn04} shows a real trend.

\subsection{Differences in the ADFs of Milky Way and M31}
 
The Milky Way and M31 are two large spiral galaxies of similar size and mass but with significantly different metallicities.  MDFs for different regions in the halo of M31 have been derived by a number of authors \citep{durrell01,reitzel02,bellazzini03,durrell04}. The MDF peak of the M31 halo is located between  \FeH $\simeq -0.4$ and $-0.7$, about 1 dex more metal rich than that of the Milky Way halo (the range in different MDF peaks may indicate that observations probe regions with different substructure).  

Our own results show that halos with the same mass and roughly the same stellar luminosity  (eg. halos H1 and H2) can have significantly different MDFs at present time due to differences in their mass accretion histories.  The \FeH peak of halo H2 is more metal rich (\FeH $\simeq -0.6$) than that of halo H1 (\FeH $\simeq -1$). As discussed before, both halos form at about the same time (see Figure \ref{fig:mass_accr}), however about $50\%$ of the stellar mass of the inner H2 halo originates in one massive satellite ($M_{*} \sim 10^{9} M_{\odot}$), whereas many smaller satellites make up the inner H1 halo (see Figure \ref{fig:fhalo_mstar}). The more massive satellite in the H2 run has enriched the stellar halo with more metal rich stars. Therefore it is possible that M31 may have had one or a few more massive mergers than the Galaxy in the first few Gyr of its formation. Alternatively, a massive satellite accreted later and on very eccentric orbit (so as to decay rapidly) could have enriched the inner halo with more Fe.  Indeed, such an accretion event seems to have been detected as a metal-rich, giant stellar stream \citep{ibata01,ferguson02}. 

Moreover, the Milky Way MDF peak is $\sim 0.5$~dex more metal poor than the most metal poor of the four simulated halos analyzed here (halo H1), suggesting that the Galaxy could have formed mainly through accretion of smaller ($M_{*} < 10^9 \, M_{\odot}$) satellites. We note that the scatter at the high end of the mass spectrum can result in significant shifts in the MDF peaks $-$ Figure \ref{fig:mdf1_halo_sats} shows variations up to about $0.5$~dex among halos. Interestingly, the only massive satellite known to be in an ongoing process of disruption, the Sagittarius galaxy, is expected to add a signficant stellar mass enriched in  \FeH to the halo in about $1-2$~Gyr, and therefore to shift the MDF peak toward the more metal rich end.

\section{Conclusions}

We have investigated the nature of the progenitors of the stellar halo for a set of  Milky Way-type galaxies and studied the chemical enrichment patterns in the context of the $\Lambda$CDM model. The main results can be summarized as follows:

$-$  In the hierarchical scenario stellar halos of Milky Way-type galaxies form inside-out. The local halo (inner $R<20$~kpc) assembles rapidly, with most of its mass already in place more than $8-9$~Gyr ago. Satellites accreted more than $9$~Gyr ago are completely disrupted by tidal forces. Satellites surviving today were accreted within the past few Gyr and fell in less radial orbits. A significant portion ($50 - 80\%$) of present day stellar halos is likely to have been originated in just a few massive satellites of $M_{*} \sim  10^{8} - 10^{10} \, M_{\odot}$. 

$-$ Our model retrieves the chemical abundance patterns observed in the Galaxy \cite[eg.][]{venn04}. Our results show that differences between the accretion times and masses of surviving  satellites versus the main progenitors of the local halo are a critical factor in determining the chemical abundance patterns. Surviving satellites have lower \alphaFe than the progenitors of the inner halo because they have been accreted later and Type Ia supernovae had enough time to enrich them with Fe. In contrast, the original building blocks were accreted fast (within the first few Gyr of the formation Galaxy) and contained stars enriched mainly by Type II supernovae. The local halo contains a significant population characterized by low \FeH and high \alphaFe ratios, originating in a few early accreted high mass systems. 

$-$ We also investigated the abundance distribution functions of both \FeH and \alphaFe ratios for stars in the halo and in surviving satellites and showed that the shapes and peaks of the distribution functions are directly related to the accretion history of the galaxies. We found a consistent shift between the peaks of the \alphaFe distribution functions of stars in surviving satellites and those in the local halo. Surviving satellites have an \alphaFe distribution peak near solar values, whereas that of the inner halo is typically around $0.1-0.2$~dex.

We conclude that the difference in chemical abundance patterns in local halo versus surviving satellites arises naturally from the predictions of the hierarchical structure formation in a $\Lambda$CDM  Universe.

\vskip0.1in
 \noindent  
A.F. would like to thank  Falk Herwig, Kim Venn and Ian McCarthy for useful discussions and suggestions. A.F and K.V.J's contributions were supported through NASA grant NAG5-9064 and NSF CAREER award AST-0133617.

\clearpage
\begin{deluxetable}{llll}
\tablewidth{450pt} 
{\label{table:halos}}\tablecaption{Properties of the four simulated halos and their satellites}
\tablecolumns{5}
\tablenum{1}
\tablehead{
\colhead{HALO} & {Stellar Halo Luminosity} & {No. of satellites} & No. of surviving \\
  & {($10^{9} \, L_{\odot}$)} & (dark matter/baryonic) & baryonic satellites
}
\startdata
H1   &  1.75 &  391 / 113 & 18 \\
H2   &  1.64 &  373 / 102 &  6 \\
H3   &  1.40 &  322 / 104 & 16 \\
H4   &  1.90 &  347 / 97 & 8 \\
\enddata
\end{deluxetable}

\end{document}